# Large current modulation and tunneling magnetoresistance change by a side-gate electric field in a GaMnAs-based vertical spin metal-oxide-semiconductor field-effect transistor

Toshiki Kanaki[1,a)], Hiroki Yamasaki[1], Tomohiro Koyama[2], Daichi Chiba[2], Shinobu Ohya[1,3,4,b)] & Masaaki Tanaka[1,3,c)]

[1]*Department of Electrical Engineering and Information Systems, The University of Tokyo, 7-3-1 Hongo, Bunkyoku, Tokyo 113-8656, Japan*

[2]*Department of Applied Physics, The University of Tokyo, 7-3-1 Hongo, Bunkyoku, Tokyo 113-8656, Japan*

[3]*Center for Spintronics Research Network (CSRN), Graduate School of Engineering, The University of Tokyo, 7-3-1 Hongo, Bunkyo-ku, Tokyo 113-8656, Japan*

[4]*Institute of Engineering Innovation, Graduate School of Engineering, The University of Tokyo, 7-3-1 Hongo, Bunkyo-ku, Tokyo 113-8656, Japan*

A vertical spin metal-oxide-semiconductor field-effect transistor (spin MOSFET) is a promising low-power device for the post scaling era. Here, using a ferromagnetic-semiconductor GaMnAs-based vertical spin MOSFET with a GaAs channel layer, we demonstrate a large drain-source current $I_{DS}$ modulation by a gate-source voltage $V_{GS}$ with a modulation ratio up to 130%, which is the largest value that has ever been reported for vertical spin field-effect transistors thus far. We find that the electric field effect on *indirect* tunneling via defect states in the GaAs channel layer is responsible for the large $I_{DS}$ modulation. This device shows a tunneling magnetoresistance (TMR) ratio up to ~7%, which is larger than that of the planar-type spin MOSFETs, indicating that $I_{DS}$ can be controlled by the magnetization configuration. Furthermore, we find that the TMR ratio can be modulated by $V_{GS}$. This result mainly originates from the electric field modulation of the magnetic anisotropy of the GaMnAs ferromagnetic electrodes as well as the potential modulation of the nonmagnetic semiconductor GaAs channel layer. Our findings provide important progress towards high-performance vertical spin MOSFETs.

[a] Corresponding author: kanaki@cryst.t.u-tokyo.ac.jp

[b] Corresponding author: ohya@cryst.t.u-tokyo.ac.jp

[c] Corresponding author: masaaki@ee.t.u-tokyo.ac.jp

Reducing the power consumption in integrated circuits is an important issue that we have to tackle in the 21st century. Making volatile components non-volatile is one of the most promising approaches to this issue; various non-volatile technologies, such as reconfigurable logic circuits[1], non-volatile power gating[2] and magnetic random access memory, are applicable to low-power-consumption electronics. A spin metal-oxide-semiconductor field-effect transistor (spin MOSFET)[3,4], in which source and drain electrodes are ferromagnetic materials, is a key component for realizing those applications because of its unique output characteristics and its compatibility with existing semiconductor technologies. For practical operation of spin MOSFETs, both large current modulation by applying a gate electric field and large magnetoresistance (MR) by magnetization reversal are required. At present, lateral and vertical types of spin MOSFETs have been proposed. The lateral spin MOSFETs, in which a current flows parallel to the substrate plane and is controlled by a gate electric field applied from the top of the channel, have a large current modulation capability ($5 \times 10^6$%[5], 400%[6] and $10^7$%[7]); however, the problem is the small MR ratio (0.1%[5], 0.005%[6] and 0.027%[7]). Meanwhile, a *vertical* spin field-effect transistor (FET)[8–13], in which a current flows perpendicular to the film plane and is controlled by a gate voltage applied from the *side surface* of the channel, is promising for large MR. Previously, we reported ferromagnetic-semiconductor GaMnAs-based vertical spin FETs that exhibit large MR ratios (60%[14] and 5%[15]). A GaMnAs-based heterostructure is one of the most ideal material systems, because we can obtain high-quality single-crystalline GaMnAs / nonmagnetic semiconductor (GaAs) / GaMnAs trilayers and thus can suppress spin relaxation at the interfaces[16–19]. However, in those vertical spin FETs, the current modulation ratio by the gate voltage was small (0.5%[14] and 20%[15]). Thus, vertical spin

FETs with a large current modulation are strongly required. In addition, to further improve the performance of the vertical spin FETs, we need more profound understanding of the gate electric field effect on the spin-dependent transport.

In vertical spin FETs, as shown by our electric field simulations later, the gate electric field influences the electric potential profile only within 10 nm from the side surfaces of the intermediate channel layer, which limits the current modulation. Thus, the lateral size of vertical FETs should be decreased as much as possible for obtaining high-performance vertical spin FETs. In this study, to enhance the current modulation and to understand the electric field effect on the spin-dependent transport, we reduced the lateral size (= width of the mesas as explained later) of the GaMnAs-based vertical spin MOSFET to ~500 nm. We have successfully obtained a large current modulation by the gate electric field with a modulation ratio up to 130%, which is the largest value that has been ever reported for vertical spin FETs[14,15]. Furthermore, using the electric field simulations, we find that *indirect* tunneling mainly contributes to the observed large current modulation. These new findings are important steps to further improve the performance of the vertical spin FETs.

**Results**

**Samples.** Our vertical spin MOSFET has a thin GaAs channel (9 nm) and ferromagnetic-semiconductor GaMnAs source and drain electrodes [Fig. 1(a)] (See the Methods section). To increase the current modulation, we reduced the width of the mesas down to ~500 nm. As a gate insulator, a 40-nm-thick $HfO_2$ film was used since it has a large relative permittivity, which also contributes to the increase of the current modulation. In this device, tunneling of holes occurs between the source and drain, because GaAs is a potential barrier with a height of ~100 meV for holes in the GaMnAs layers[20,21], as shown in Fig. 1(b). When the gate-source voltage $V_{GS} < 0$ V ($V_{GS} \geq 0$ V), the tunneling current flowing at the side surfaces of the mesas is increased (decreased), as shown in Fig. 1(c).

**MOSFET operation and its analyses.** To investigate the MOSFET characteristics of this device, we measured the drain-source current $I_{DS}$ as a function of the drain-source voltage $V_{DS}$ for various $V_{GS}$ [Fig. 2(a)]. Nonlinear $I_{DS}$–$V_{DS}$ characteristics were observed for each $V_{GS}$ (black curves), indicating that tunneling transport is dominant. Furthermore, $I_{DS}$ was largely controlled by $V_{GS}$. We note that the gate leakage current and electric field effect on parasitic resistances (the resistances of the top/bottom GaMnAs layers, GaAs:Be layer and Au/Cr electrodes), which may induce unintended modulation of $I_{DS}$, were negligibly small (see Supplementary Note 1). When $V_{GS} = 20$ V, the $I_{DS}$ modulation ratio by $V_{GS}$, which is defined by $[I_{DS}(V_{GS}) - I_{DS}(V_{GS} = 0 \text{ V})] / I_{DS}(V_{GS} = 0 \text{ V})$, is around –20% [see the blue points in Fig. 2(b)]. On the other hand, when $V_{GS} = -20$ V, it reached ~130% [see the red points in Fig. 2(b)]. This $I_{DS}$ modulation ratio (~130%) is the largest among the values reported for vertical spin FETs thus far[14,15].

To understand the modulation of the band alignment in detail, we measured $I_{DS}$ as a function of $V_{GS}$ at $V_{DS} = -10$ mV [Fig. 2(c)]. $I_{DS}$ normalized at $V_{GS} = 0$ V ($\gamma$) was changed from 1 to 1.28 when $V_{GS}$ was changed from 0 V to –3 V [see the right axis in Fig. 2(c)], meaning that $I_{DS}$ was increased by 28% when $V_{GS}$ was changed from 0 V to –3 V. As shown below, this large modulation of $I_{DS}$ cannot be understood by the electric field effect on *direct* tunneling. To obtain the potential distribution and to calculate $I_{DS}$ normalized at $V_{GS} = 0$ V ($\gamma_{calc}$), we performed electric field simulation varying the electric potential at the side surface of the mesa and investigated the effect of the side-gate electric field [Fig 2(e,h)] (see Supplementary Note 2). Here, we define $E_V^{(S)}$ as the valence band top energy $E_V$ at the side surface (interface between the side-gate electrode and the GaAs channel) with respect to the Fermi level $E_F$ in terms of hole energy. The potential profile of $E_V$ when $E_V^{(S)} = 0.75$ eV is shown in Fig. 2(e), which corresponds to the case of $V_{GS} = 0$ V, because $E_F$ at the side surface of the GaAs channel is pinned at the middle of the band gap[22]. With decreasing $E_V^{(S)}$ from 0.75 eV, the electric potential near the side surface of the mesa is decreased [Fig. 2(f,i)], whereas the electric potential in the inner region of the mesa (10 nm ≤ $x$) is not influenced [Fig. 2(g,j)]. As shown in Fig. 2(d), $\gamma_{calc}$ remains almost unchanged between Fig. 2(e) ($\gamma_{calc} = 1$ when $E_V^{(S)} = 0.75$ eV) and (h) ($\gamma_{calc} = 1.028$ when $E_V^{(S)} = 0.15$ eV) because GaAs is a potential barrier for holes in both cases. On the other hand, when $E_V^{(S)} < 0.15$ eV, because $E_V$ of the GaAs channel at the side surface becomes lower than $E_V$ inside the mesa, $\gamma_{calc}$ starts to increase with decreasing $E_V^{(S)}$ [Fig. 2(d)]. This feature is different from the experimental data shown in Fig. 2(c); when $V_{GS}$ is changed from 0 V to –10 V, $\gamma$ starts to increase at $V_{GS} = 0$ V. We can see the significant difference in the curve shapes of Fig. 2(c) and Fig. 2(d). This analysis indicates that the electric field effect on the *direct* tunneling current *cannot* reproduce the experimental $I_{DS}$–$V_{GS}$ characteristic. Instead, the main origin of the large modulation ratio observed in our device is the electric field effect on the *indirect* tunneling current[23].

The indirect tunneling current is probably caused by a large amount of Mn atoms (~$10^{18}$ cm$^{-3}$), which are diffused to the intermediate GaAs layer from the top and bottom GaMnAs layers and form defect states in the band gap of GaAs. Furthermore, GaAs grown at low temperature (200 °C) is known to have a large amount of arsenic antisite defects ($10^{18} - 10^{19}$ cm$^{-3}$)[24,25]. In fact, the $I_{DS}$–$V_{DS}$ characteristics of our device show a strong temperature dependence (see Supplementary Note 3), which indicates that indirect tunneling via defect states takes place. (If $I_{DS}$ were dominated only by the direct tunneling current, no temperature dependence would be observed.) Therefore, the electric field effect on indirect tunneling via defect states is the most probable origin for the large $I_{DS}$ modulation ratio.

**Tunneling magnetoresistance and its change by $V_{GS}$.** To investigate the spin-dependent transport of our device, we measured the drain-source resistance $R_{DS}$ as a function of $\mu_0 H$ applied along the [$\bar{1}10$] direction in the film plane with $V_{DS} = -5$ mV and $V_{GS} = 0$ V [Fig. 3(a)]. Here, $R_{DS}$ is defined by $V_{DS}/I_{DS}$, $\mu_0$ is the permeability of a vacuum and $H$ is an in-plane external magnetic field. In the major loop (black circles), clear tunnel magnetoresistance (TMR) was observed, indicating that $I_{DS}$ can be controlled by the magnetization configuration. The TMR ratio, which is defined by [$R_{DS}(\mu_0 H) - R_{DS}(\mu_0 H = 0$ mT$)]/R_{DS}(\mu_0 H = 0$ mT$) \times 100$ (%), reached ~7% at $\mu_0 H = 20$ mT, where $R_{DS}(\mu_0 H)$ is the drain-source resistance at $H$ in the major loop. This value is more than 70 times larger than the MR ratios obtained in the lateral spin MOSFETs[5–7]. We also observed a clear minor loop (red circles), indicating that the antiparallel magnetization configuration is stable even at $\mu_0 H = 0$ mT. (In the minor loop, $R_{DS}$ increases with increasing $\mu_0 H$ from –20 mT to 60 mT, probably because the magnetizations of the top and bottom GaMnAs layers are not completely antiparallel at the peak of $R_{DS}$ (at $\mu_0 H = -20$ mT in the major loop) and they become close to the perfect antiparallel configuration with increasing $\mu_0 H$ to 10 mT in the minor loop.)

To investigate the influence of $V_{GS}$ on the spin-dependent transport, we measured the $V_{GS}$ dependence of the TMR ratio [Fig. 3(b)]. Here, the TMR ratio corresponds to the maximum value obtained in the major loop at each $V_{GS}$. The TMR ratio tends to increase as $V_{GS}$ is changed from 0 V to -10 V. In our device, the gate electric field can modulate the electronic states of the top/bottom GaMnAs layers as well as those of the intermediate GaAs layer, both of which can modulate TMR. Applying $V_{GS}$ causes the change in the hole density of the GaMnAs layers near the side surfaces of the mesas, which can change the spin polarization and magnetic anisotropy[15]. To understand the modulation of the magnetic anisotropy by $V_{GS}$ in our device, we measured TMR applying $H$ in various in-plane directions with an angle $θ$ with respect to the [100] axis in the counterclockwise rotation when $V_{GS}$ = 0, –5 and –10 V [Fig. 3(c–e)] (see Measurements section). The observed TMR ratios showed dominant uniaxial anisotropy along the [$\bar{1}$10] direction in addition to biaxial anisotropy along the <100> directions for any $V_{GS}$ [see the four red peaks in Fig. 3(c–e)]. With changing $V_{GS}$ from 0 V to –10 V, the easy axes of our device were slightly rotated toward the [010] direction (the red-colored region is extended toward the [010] direction). Furthermore, the coercive force of the top GaMnAs layer, which has a larger coercivity than the bottom GaMnAs layer, increases as $V_{GS}$ is changed from 0 V to –10 V (the red-colored region slightly expands outward). These results indicate that the magnetic anisotropy constants are modulated by applying negative $V_{GS}$, which can also contribute to the modulation of the TMR ratio. In addition, $V_{GS}$ modifies the electric potential of the GaAs layer. As we discussed in the previous paragraph, the modulation of the indirect tunneling current via defect states is the most probable mechanism for the obtained large modulation of $I_{DS}$. The TMR induced by indirect tunneling via defect states depends on many factors such as energy levels of defect states, band width of defect states, and life time of carries and so on. The modulation of the electric potential can influence the indirect tunneling and thus TMR. Therefore, the electric field effect both on the top/bottom GaMnAs layers and on the intermediate GaAs layer contributes to the modulation of the observed TMR ratio.

Surprisingly, the $V_{GS}$ dependence of the TMR ratio shown in Fig. 3(b) is completely opposite to the one obtained in our previous study, *i.e.* the TMR ratio *decreased* as negative $V_{GS}$ is applied in our previous study[14]. This may be caused by the difference of the easy axes between the present device and the previous one (the biaxial anisotropy was dominant in our previous work) or by the different direction of an external magnetic field (along the direction with an angle 10-degree from the [100] direction toward the [1$\bar{1}$0] in our previous work).

**Summary**

In summary, we have investigated the electric field effect on the spin-dependent transport properties in a GaMnAs-based vertical spin MOSFET. We obtained a large current modulation ratio up to 130 %, which is the largest value that has ever been reported thus far for the vertical spin FETs[14,15]. By comparing the experimental data with the calculated results, we concluded that this large $I_{DS}$ modulation does not originate from the modulation of direct tunneling between the source and the drain but from the modulation of the indirect tunneling current via defect states in the intermediate GaAs layer. The TMR ratio tends to increase as negative $V_{GS}$ is applied, which is attributed to the electric field effect both on the top/bottom GaMnAs layers and on the intermediate GaAs layer. These results provide an important insight into the device physics for realizing high-performance vertical spin MOSFETs.

**Methods**

**Growth.** The heterostructure composed of, from the top to the bottom, $Ga_{0.94}Mn_{0.06}As$ (10 nm) / GaAs (9 nm) / $Ga_{0.94}Mn_{0.06}As$ (3.2 nm) / GaAs:Be (50 nm, hole concentration $p = 5 \times 10^{18}$ cm$^{-3}$) on a $p^+$-GaAs (001) substrate by low-temperature molecular beam epitaxy. The growth temperatures of the top $Ga_{0.94}Mn_{0.06}As$, GaAs, bottom $Ga_{0.94}Mn_{0.06}As$ and GaAs:Be layers were 195 °C, 180 °C, 200 °C and 520 °C, respectively.

**Process.** After the growth, we partially etched the grown films and buried the etched area with a 100-nm-thick $SiO_2$ layer for the separation of the drain electrode and the substrate. Then, a comb-shaped Au (40 nm) / Cr (5 nm) layer, whose width of the comb teeth is ~500 nm and length of them is 50 μm, was formed by electron-beam lithography and a lift-off technique. We chemically etched the area that is not covered by the Au/Cr layer and then the magnetic tunnel junctions only beneath the comb teeth area of the Au/Cr drain electrode remained after the etching. We formed a 40-nm-thick $HfO_2$ film as a gate insulator using atomic layer deposition at a substrate temperature of 150 °C and deposited a gate electrode composed of Au (50 nm) / Cr (5 nm) by electron-beam deposition.

**Measurements.** After the device was bonded with Au wires and indium solder, we measured the spin-dependent transport properties of our spin MOSFET with varying $V_{GS}$ and $H$ at 3.8 K. To measure the $\theta$ dependence of TMR, we applied a strong magnetic field of 1 T in the opposite direction of $\theta$ to align the magnetization directions and we decreased $H$ to zero. Then, we started to measure $R_{DS}$ while increasing $H$ from zero in the direction of $\theta$. The measurements were performed at every 10° step of $\theta$.

Acknowledgements

This work was partly supported by Grants-in-Aid for Scientific Research (No. 26249039, No. 16H02095), CREST program of Japan Science and Technology Agency, and Spintronics Research Network of Japan (Spin-RNJ). T. Kanaki was supported by JSPS through the program for leading graduate schools (MERIT). T. Kanaki thanks the JSPS Research Fellowship Program for Young Scientists.

Author contributions

Sample preparation: T. Kanaki, H. Y., T. Koyama., and D. C.; measurements: T. Kanaki and H. Y.; data analysis: T. Kanaki and H. Y.; writing and project planning: T. Kanaki, S. O., and M. T.

Figure captions

FIG. 1. (Color online) (a) Schematic illustration of the vertical spin MOSFET investigated in this study. The backside of the substrate is the source electrode, the comb shaped Au/Cr layer is the drain electrode and the Au/Cr layer above the $HfO_2$ layer is the gate electrode. (b)(c) Schematic device operation of our vertical spin MOSFET when a gate voltage $V_{GS}$ is not applied (b) and when a negative gate voltage is applied (c). The orange arrows represent a drain-source current $I_{DS}$.

FIG. 2. (Color online) (a) Drain-source current $I_{DS}$ as a function of the drain-source voltage $V_{DS}$ with the gate-source voltage $V_{GS}$ ranging from –20 V to 20 V with a step of 5 V at 3.8 K. (b) $I_{DS}$ modulation ratio as a function of $V_{DS}$ with various $V_{GS}$ at 3.8 K. (c) Drain-source current ($-I_{DS}$) (left axis) and the $I_{DS}$ value normalized at $V_{GS}$ = 0 V ($\gamma$) (right axis) as a function of $V_{GS}$ with $V_{DS}$ = –10 mV at 3.8 K. (d) Calculated $I_{DS}$ normalized by the one at $V_{GS}$ = 0 V ($\gamma_{calc}$) as a function of $E_V^{(S)}$. (e)(h) Calculated valence band top energy $E_V$ with respect to the Fermi level when $E_V^{(S)}$ = 0.75 eV (e) and $E_V^{(S)}$ = 0.15 eV (h). Here, the Fermi level corresponds to 0 eV. The vertical axis expresses the hole energy. The inset in (e) and (h) shows the structure used in our calculation. Here, the $x$ axis represents the distance from the side surface of the mesa and the $y$ axis denotes the distance from the interface between the bottom GaMnAs layer and the intermediate GaAs layer. The calculation was performed in the region surrounded by the dashed line. In (e,h), only the region of 0 nm $\leq x \leq$ 15 nm is shown because it is sufficient to see how the gate electric field influences the electric potential in the GaAs layer. (f,g) $E_V$ vs. $y$ at $x$ = 1 nm (f) and 15 nm (g) when $E_V^{(S)}$ = 0.75 eV. (i,j) $E_V$ vs. $y$ at $x$ = 1 nm (i) and 15 nm (j) when $E_V^{(S)}$ = 0.15 eV.

FIG. 3. (Color online) (a) Drain-source resistance $R_{DS}$ as a function of the in-plane external magnetic field $\mu_0 H$ applied along the [$\bar{1}$10] direction at 3.8 K. Here, the drain-source voltage $V_{DS}$ was –5 mV and the gate-source voltage $V_{GS}$ was 0 V. The black circles correspond to the major loop and the red circles correspond to the minor loop. The black (red) arrows are the sweep directions in the major (minor) loop. The magnetization states in the major loop are indicated by the white arrows above the graph. (b) TMR ratio as a function of the gate-source voltage $V_{GS}$ at 3.8 K. Here, the drain-source voltage $V_{DS}$ was fixed at –5 mV and the external magnetic field $H$ was applied along the [$\bar{1}$10] direction. The TMR ratio is the maximum value obtained in the major loop at each $V_{GS}$. (c–e) Magnetic-field-direction dependences of the TMR ratios at $V_{DS}$ = –10 mV with $V_{GS}$ = 0 V (c), –5 V (d) and –10 V (e).

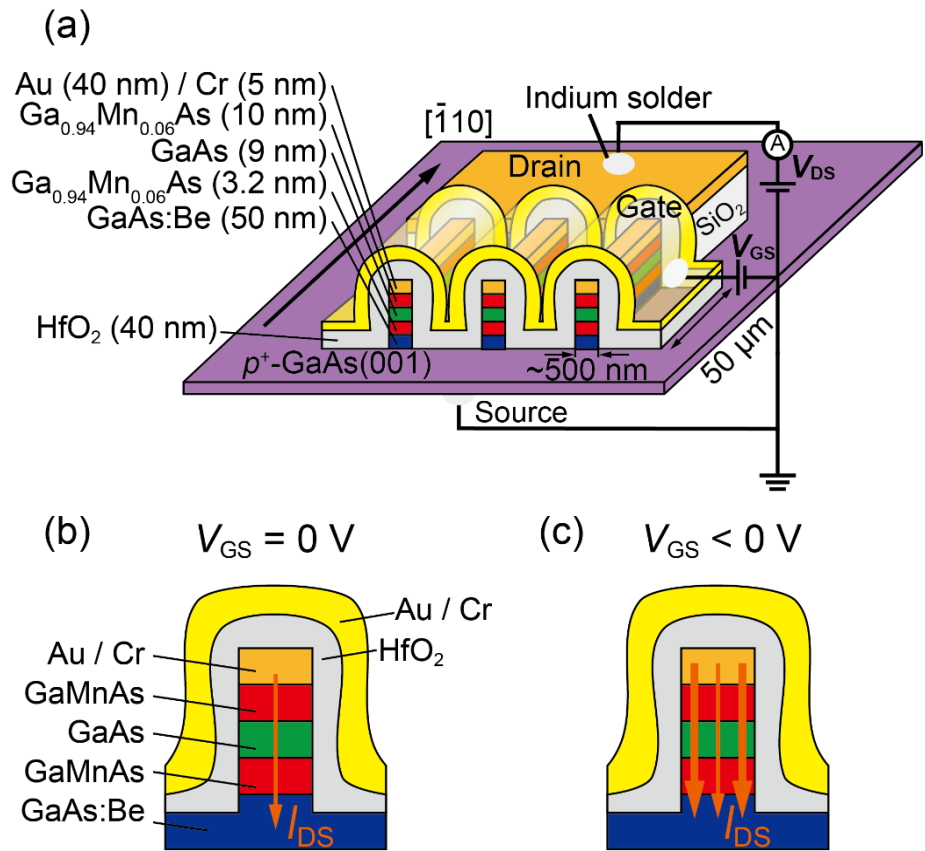

Fig. 1

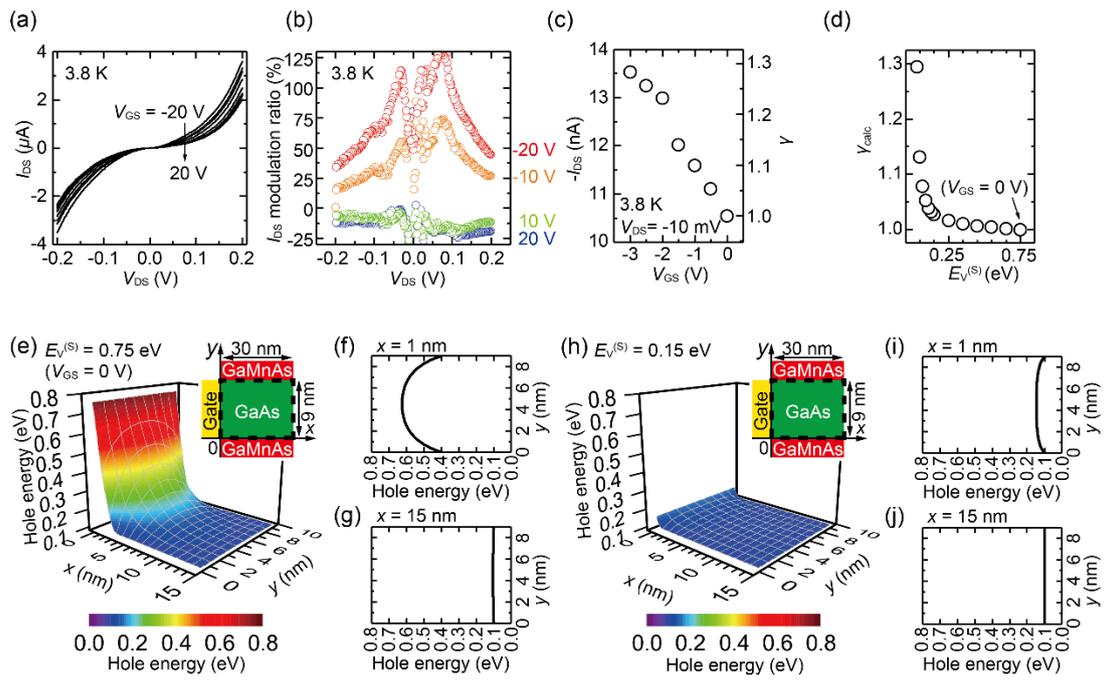

Fig. 2

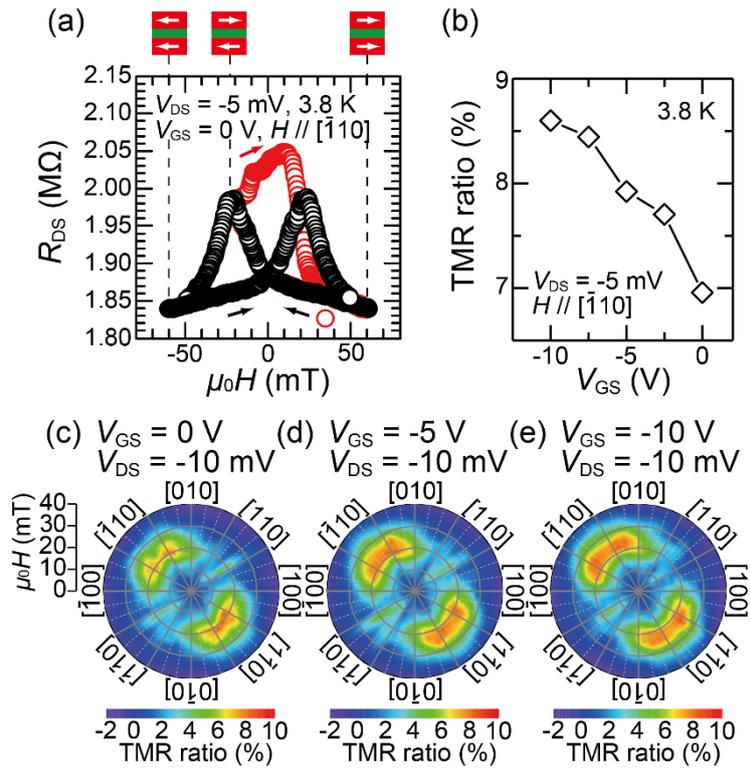

Fig. 3

Supplementary information

Large current modulation and tunneling magnetoresistance change by a side-gate electric field in a GaMnAs-based vertical spin metal-oxide-semiconductor field-effect transistor

Toshiki Kanaki, Hiroki Yamasaki, Tomohiro Koyama, Daichi Chiba, Shinobu Ohya and Masaaki Tanaka

**Supplementary Note 1. Influence of the gate leakage current and electric field effect on the parasitic resistances**

The observed gate-source leakage current ranged from -400 pA to 200 pA. It is much smaller than the experimentally observed drain-source current ($I_{DS}$) modulation (~0.36 µA at the drain-source voltage ($V_{DS}$) of 80 mV and the gate-source voltage ($V_{GS}$) of -20 V). Thus, the gate leakage current is negligibly small. The influence of the gate electric field on the parasitic resistances is also negligibly small. The resistances of the top GaMnAs layer, bottom GaMnAs layer, GaAs:Be layer and Au/Cr layers are estimated to be 1.6 mΩ, 0.5 mΩ, 20 mΩ and 10 µΩ, respectively. Here, the resistances of the top and bottom GaMnAs layers, GaAs:Be layer and Au/Cr layers are obtained using the area of the mesas, the thickness of each layer and the resistivity of GaMnAs, GaAs:Be and Au/Cr. The resistivity of the GaMnAs, GaAs:Be and Au/Cr layers was measured to be 4 mΩ.cm, 10 mΩ.cm and 5.8 µΩ.cm, respectively, using Hall-bar structures. The modulation of resistance in the GaAs:Be layer by a gate electric field was also measured to be ~3% at 3.8 K using a Hall-bar structure with a gate electrode. Because the carrier concentration of the GaMnAs layers and Au/Cr layers are much larger than that of the GaAs:Be layer, the modulation of resistance in the top and bottom GaMnAs layers and Au/Cr layers is expected to be much less than 3%. Considering that the drain-source resistance changes from 80 kΩ (at $V_{DS}$ = 200 mV) to 1.5 MΩ (at $V_{DS}$ = 5 mV) at 3.8 K (see the purple curve in Fig. S2), the modulation of the parasitic resistances is (1.6 mΩ + 0.5 mΩ + 20 mΩ + 10 µΩ)×3% / 80 kΩ ~0.8×10$^{-6}$% at most, which is much smaller than the experimentally obtained value (~130% at $V_{DS}$ = 80 mV and $V_{GS}$ = -20 V). Thus, the influence of the gate leakage current and electric field effect on the parasitic resistances is negligibly small.

**Supplementary Note 2. Details of the calculation of the electric potential profile and the normalized drain-source current in our vertical spin MOSFET**

The device structure used in our simulation is shown in Fig. S1. To calculate the electric potential profile with various gate voltages, we solved the following two-dimensional Poisson equation.

$$\frac{\partial^2 E_V}{\partial x^2} + \frac{\partial E_V^2}{\partial y^2} = \frac{\rho}{\epsilon}. \tag{S1}$$

Here, $E_V$ is the valence band top energy of GaAs in terms of hole energy, $\rho$ is the charge density and $\varepsilon$ is the dielectric constant of GaAs, respectively. Because all the experimental results presented in the main text were obtained at 3.8 K, activation of donors/acceptors and thermally excited carriers can be neglected in GaAs. Thus, we set the charge density at 0 in GaAs, meaning that the right side of Equation (S1) is 0.

Considering that the potential barrier height of GaAs is ~0.1 eV for holes in the GaMnAs layers, Equation (S1) is solved under the following boundary conditions.

- At $x = 0$, $E_V - E_F = E_V^{(S)}$.
- At $x = W$, $E_V - E_F = 0.1$ eV.[1,2]

Here, $E_F$ is the Fermi level, $E_V^{(S)}$ is the valence band top energy with respect to $E_F$ at $x = 0$ (the interface between the side-gate electrode and GaAs), $L$ is the channel length and $W$ is the width of the calculated region [see Fig. S1].

As the boundary condition at the interfaces of GaMnAs/GaAs ($y = 0$ and $L$), we used the $E_V$ profile obtained for GaMnAs; in the surface depletion region of GaMnAs ($0 \leq x \leq W_D$), where $W_D$ is the width of the depletion layer of GaMnAs, $E_V$ has a parabolic form that satisfies $E_V - E_F = E_V^{(S)}$ at $x = 0$. When $W_D \leq x$, $E_V - E_F$ was approximated to be 0.1 eV, which is the same as the barrier height of GaAs for holes, in GaMnAs. Considering the above conditions, the energy profile at $y = 0$ and $L$ can be expressed by the following equation.

$$E_V - E_F = \begin{cases} \left(E_V^{(S)} - 0.1\right)\left(1 - \frac{x}{W_D}\right)^2 + 0.1 & (0 \leq x \leq W_D) \\ 0.1 & (W_D \leq x) \end{cases}. \tag{S2}$$

The depletion layer width $W_D$ is expressed by

$$W_D = \sqrt{\frac{2\epsilon |E_V^{(S)}|}{eN_A}}. \quad (S3)$$

Here, $N_A$ is an acceptor concentration of GaMnAs and we set $N_A$ at $10^{20}$ cm$^{-3}$. To introduce the effect of the gate electric field, we changed $E_V^{(S)}$. When a gate voltage is not applied, $E_V^{(S)} = E_g/2$, where $E_g$ is the band gap of GaAs, because of the Fermi level pinning at the surface of GaAs. We set $W = 30$ nm, which is larger than the depletion layer width in the GaAs layer (2–15 nm) [see Fig. 2(e,h)]. The parameters used in the potential calculation are summarized in Table S1. Equation (S1) was solved using Jacobi's iterative method so that the potential difference between the present step and the previous step at all $(x, y)$ becomes less than $10^{-9}$ eV. Each mesh is a rectangle with a width in the $x$ direction ($\Delta x$) of 0.1 nm and a width in the $y$ direction ($\Delta y$) of 0.1 nm.

The drain-source current $I_{DS}$ at $E_V^{(S)}$ can be expressed by the following equation.

$$I_{DS} = 2 \int_{W_D}^{\frac{W_{mesa}}{2}} J_{E_V^{(S)}}(x)dx \times L_{mesa} \times N_{mesa}. \quad (S4)$$

Here, $W_{mesa}$ is the width of our mesa (500 nm), $J_{E_V^{(S)}}(x)$ is the current density at $x$ and $E_V^{(S)}$, $L_{mesa}$ is the length of our mesa (50 μm) and $N_{mesa}$ is the number of the mesas (10). When the applied voltage between the source and the drain $|V|$ is much smaller than the barrier height, the carrier energy $E$ dependence of tunneling probability can be neglected. Thus, $J_{E_V^{(S)}}(x)$ can be described by

$$\begin{aligned} J_{E_V^{(S)}}(x) &\propto \int_{-eV}^{0} D_{top}(E) D_{bot}(E+eV) T_{E_V^{(S)}}(x) \, dE \\ &= T_{E_V^{(S)}}(x) \times \int_{-eV}^{0} D_{top}(E) D_{bot}(E+eV) \, dE. \end{aligned} \quad (S5)$$

Here, $D_{top}$, $D_{bot}$ and $T_{E_V^{(S)}}(x)$ are the density of states in the top GaMnAs layer, the density of states in the bottom GaMnAs layer and the tunneling probability at $x$ and $E_V^{(S)}$, respectively.

Using equation (S4) and (S5), we can obtain the following relationship between $I_{DS}$ and $T_{E_V^{(S)}}(x)$.

$$I_{DS} \propto 2 \int_{W_D}^{\frac{W_{mesa}}{2}} T_{E_V^{(S)}}(x) dx. \tag{S6}$$

Note that we consider that $D_{top}$ and $D_{bot}$ do not depend on $E_V^{(S)}$. Therefore, only $T_{E_V^{(S)}}(x)$ is dependent on $E_V^{(S)}$.

$T_{E_V^{(S)}}(x)$ can be calculated using the Wentzel-Kramers-Brilluion approximation. The right side of Equation (S6) can be calculated as follows.

$$\begin{aligned} 2 &\int_{W_D}^{\frac{W_{mesa}}{2}} T_{E_V^{(S)}}(x) dx \\ &= 2 \int_{W_D}^{W} \exp\left(-2 \int_0^L \frac{\sqrt{2m_h \left(E_{V,E_V^{(S)}}(x,y) - E_F\right)}}{\hbar} dy\right) dx \\ &+ 2 \left(\frac{W_{mesa}}{2} - W\right) \\ &\times \exp\left(-2 \int_0^L \frac{\sqrt{2m_h \left(E_{V,E_V^{(S)}}(x=W,y) - E_F\right)}}{\hbar} dy\right). \end{aligned} \tag{S7}$$

Here, $E_{V,E_V^{(S)}}(x,y)$ is the valence band top energy at $(x, y)$ and $E_V^{(S)}$, and $\hbar$ is the reduced Planck constant.

Thus, the calculated $I_{DS}$ normalized at $V_{GS} = 0$ V, corresponding to $E_V^{(S)} = E_g/2$, $\gamma_{calc}$ can be described as follows.

$$\gamma_{\text{calc}} = \int_{W_{\text{D}}}^{\frac{W_{\text{mesa}}}{2}} T_{E_{\text{V}}^{(\text{S})}}(x)dx \bigg/ \int_{W_{\text{D}}}^{\frac{W_{\text{mesa}}}{2}} T_{E_{\text{V}}^{(\text{S})} = \frac{E_{\text{g}}}{2}}(x)dx. \tag{S8}$$

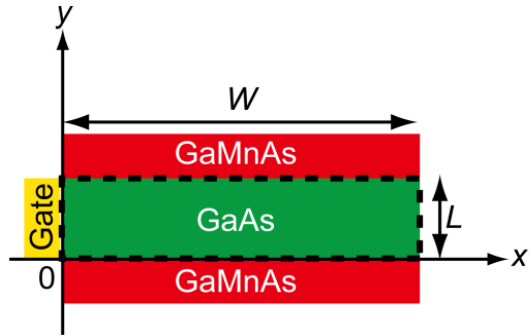

FIG. S1. Schematic illustration of the device structure used in our electric potential calculation. Here, $L$ is the GaAs-channel length (9 nm) and $W$ is the width of the calculated area surrounded by the dashed lines. In our calculation, $W$ was set at 30 nm, which is large enough to see how the gate electric field influences the electric potential profiles.

TABLE S1. Parameters used in the calculation of the electric potential profiles and $\gamma_{\text{calc}}$.

| Parameters (unit) | Values |
| --- | --- |
| $L$ (nm) | 9 |
| $W$ (nm) | 30 |
| $\Delta x$ (nm) | 0.1 |
| $\Delta y$ (nm) | 0.1 |
| $m_{\text{h}}$ (kg) | $0.45 m_0$[1,2] |
| $E_{\text{g}}$ (eV) | 1.51914 |

**Supplementary Note 3. Temperature dependence of the drain-source resistance $R_{DS}$.**

We show $R_{DS}$ *vs.* $V_{DS}$ at various temperatures in Supplementary Fig. S2. If $I_{DS}$ were dominated by direct tunneling, $R_{DS}$ *vs.* $V_{DS}$ characteristics would not depend on the temperature. In our vertical spin MOSFET, when the temperature was changed from 300 K to 3.8 K, $R_{DS}$ was changed from 2 kΩ to 1.2 MΩ at $V_{DS}$ = -10 mV. The strong temperature dependence of $R_{DS}$ is experimental evidence of the indirect tunneling via defect states.

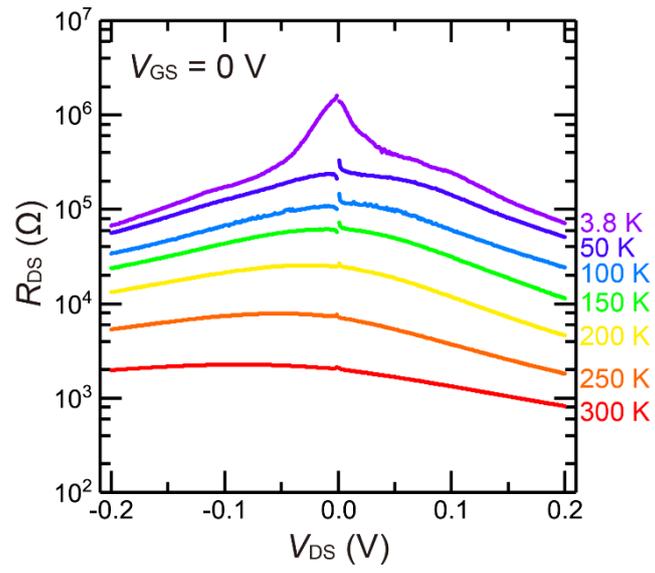

FIG. S2. Drain-source resistance $R_{DS}$ as a function of $V_{DS}$ at various temperatures in our vertical spin MOSFET. Here, $V_{GS} = 0$ V.